\documentclass[allclo]{FBSart}
\usepackage{amsfonts}
\usepackage{amssymb}
\usepackage{epsfig}

\title{Recent Progress in Effective Field Theory in the One-Nucleon Sector}
\author{S. Scherer\thanks{\textit{E-mail address:}
scherer@kph.uni-mainz.de}} \institute{Institut f\"ur Kernphysik,
Johannes Gutenberg-Universit\"at Mainz, J.~J.~Becher-Weg 45, D-55099
Mainz, Germany}

\runningauthor{S.\ Scherer} \runningtitle{Recent Progress in
Effective Field Theory in the Single-Nucleon Sector} 

\begin{document}

\maketitle
\begin{abstract}
   Chiral perturbation theory (ChPT) is the effective field theory of the strong
interactions at low energies.
   We will address the issue of a consistent power counting
scheme in a manifestly Lorentz-invariant formulation of baryon ChPT.
   As applications we show how the inclusion of vector and axial-vector mesons
in the calculation of the nucleon electromagnetic and axial form
factors, respectively, lead to an improved description of the
empirical data.
   Finally, we will outline a systematic implementation of the $\Delta(1232)$
resonance into the effective field theory program.

\end{abstract}

\section{Introduction}

   Effective field theory (EFT) is a powerful tool for describing the
strong interactions at low energies.
   The EFT of the interactions among the Goldstone bosons
   of spontaneous chiral symmetry breaking in QCD is
(mesonic) chiral perturbation theory (ChPT)
\cite{Weinberg:1978kz,Gasser:1983yg} (see, e.g.,
Ref.~\cite{Scherer:2002tk} for a pedagogical introduction).
    Besides the most general Lagrangian a successful EFT program requires a consistent
power counting scheme to assess the importance of a given
renormalized diagram.
   In the following we will outline some recent
developments in devising a renormalization scheme leading to a
simple and consistent power counting for manifestly
Lorentz-invariant baryon ChPT \cite{Gasser:1987rb}.
   The approach allows for
both the inclusion of further degrees of freedom beyond pions and
nucleons \cite{Fuchs:2003sh,Hacker:2005fh} and the application to
higher-loop calculations \cite{Schindler:2003je,Schindler:2006ha}.

\section{Renormalization and Power Counting}
   The Lagrangian of the most general chirally invariant
interaction of pions and nucleons in the one-nucleon sector is
organized in a derivative and quark-mass expansion
\cite{Weinberg:1978kz,Gasser:1983yg,Scherer:2002tk,Gasser:1987rb}:
\begin{displaymath}
{\cal L}_{\rm eff}={\cal L}_{\pi}+{\cal L}_{\pi N}={\cal
L}_\pi^{(2)}+{\cal L}_\pi^{(4)}+\cdots+{\cal L}_{\pi N}^{(1)}+{\cal
L}_{\pi N}^{(2)}+\cdots.
\end{displaymath}
   The aim is to devise a renormalization procedure generating, after
renormalization, the following power counting:
   a loop integration in $n$ dimensions counts as $q^n$,
pion and fermion propagators count as $q^{-2}$ and $q^{-1}$,
respectively, vertices derived from ${\cal L}_\pi^{2k}$ and ${\cal
L}_{\pi N}^{(k)}$ count as $q^{2k}$ and $q^k$, respectively.
   Here, $q$ generically denotes a small expansion parameter such as,
e.g., the pion mass.

   In order to illustrate the issue of power counting, we consider as an example
the one-loop integral (in the chiral limit)
\begin{displaymath}
H(p^2,m^2;n)= \int \frac{d^n k}{(2\pi)^n}
\frac{i}{[(k-p)^2-m^2+i0^+][k^2+i0^+]},
\end{displaymath}
where $\Delta=(p^2-m^2)/m^2={\cal O}(q)$ is a small quantity.
   Applying the dimensional counting analysis of Ref.~\cite{Gegelia:1994zz},
the result of the integration is of the form
\begin{displaymath}
H\sim F(n,\Delta)+\Delta^{n-3}G(n,\Delta),
\end{displaymath}
where $F$ and $G$ are hypergeometric functions which can be expanded
in $\Delta$ for any $n$.
   In the present case, we want the renormalized integral to be of the order
$D=n-1-2=n-3$.

   The infrared regularization of Becher and Leutwyler
\cite{Becher:1999he} makes use of the Feynman parametrization
\begin{displaymath}
{1\over ab}=\int_0^1 {dz\over [az+b(1-z)]^2}
\end{displaymath}
with $a=(k-p)^2-m^2+i0^+$ and $b=k^2+i0^+$.
   The resulting integral over the Feynman parameter $z$ is then rewritten as
\begin{eqnarray*}
H=\int_0^1 dz \cdots &=& \int_0^\infty dz \cdots
- \int_1^\infty dz \cdots,\\
\end{eqnarray*}
where the first, so-called infrared (singular) integral satisfies
the power counting, while the remainder violates power counting but
turns out to be regular and can thus be absorbed in counterterms.

   The central idea of the extended on-mass-shell (EOMS)
scheme \cite{Fuchs:2003qc} consists of performing additional
subtractions beyond the $\widetilde{\rm MS}$ scheme of Ref.\
\cite{Gasser:1987rb}.
   Since the terms violating the power counting are analytic in small
quantities, they can be absorbed by counterterm contributions.
 To that end one first expands the integrand in
small quantities and subtracts those (integrated) terms whose order
is smaller than suggested by the power counting.
   The corresponding subtraction term reads
\begin{displaymath}
H^{\rm subtr}=\int \frac{d^n k}{(2\pi)^n}\left.
\frac{i}{[k^2-2p\cdot k +i0^+][k^2+i0^+]}\right|_{p^2=m^2}
\end{displaymath}
and the renormalized integral is written as $ H^R=H-H^{\rm
subtr}={\cal O}(q) $ as $n\to 4$.

   In Ref.\ \cite{Schindler:2003xv} the IR regularization of
Becher and Leutwyler was reformulated in a form analogous to the
EOMS renormalization scheme.
   Within this (new) formulation the subtraction terms are found by
expanding the integrands of loop integrals in powers of small
parameters (small masses and Lorentz-invariant combinations of
external momenta and large masses) and subsequently exchanging the
order of integration and summation.
   The new formulation of IR regularization can be applied to
diagrams with an arbitrary number of propagators with various masses
(e.g., resonances) and/or diagrams with several fermion lines as
well as to multi-loop diagrams.

\section{Applications}

\subsection{Electromagnetic Form Factors}
\begin{figure}[hbt]
\begin{center}
\epsfig{file=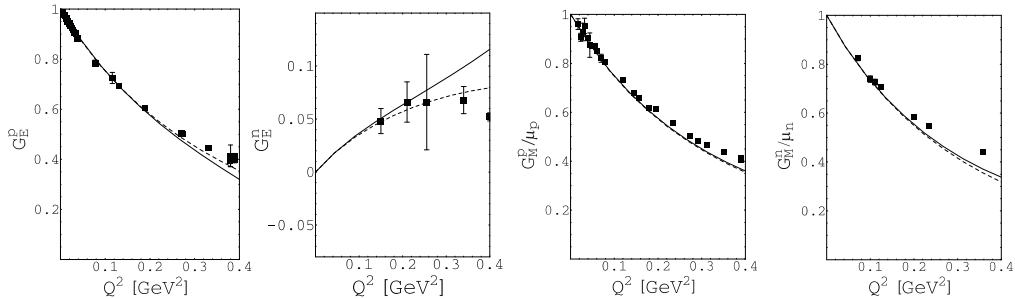,width=\textwidth}
\end{center}
\caption{\label{G_neu} The Sachs form factors of the nucleon in
manifestly Lorentz-invariant chiral perturbation theory at ${\cal
O}(q^4)$ including vector mesons as explicit degrees of freedom.
Full lines: results in the extended on-mass-shell scheme; dashed
lines: results in infrared regularization.}
\end{figure}

   The nucleon matrix element of the electromagnetic current operator
$J^\mu(x)$ can be parameterized in terms of two form factors,
\begin{displaymath}
\langle N(p')|J^{\mu}(0)|N(p)\rangle=
\bar{u}(p')\left[F_1^N(Q^2)\gamma^\mu
+i\frac{\sigma^{\mu\nu}q_\nu}{2m_p}F_2^N(Q^2) \right]u(p),\quad
N=p,n,
\end{displaymath}
   where $q=p'-p$, $Q^2=-q^2$, and $m_p$ is the proton mass.
   At $Q^2=0$, the so-called Dirac and Pauli form factors $F_1$ and
$F_2$ reduce to the charge and anomalous magnetic moment in units of
the elementary charge and the nuclear magneton $e/(2m_p)$,
respectively:
$F_1^{p}(0)=1$, $F_1^{n}(0)=0$, $F_2^{p}(0)=1.793$, and
$F_2^{n}(0)=-1.913$.
   The Sachs form factors $G_E$ and $G_M$ are linear combinations of $F_1$ and
$F_2$,
\begin{displaymath}
G_E^N(Q^2)=F_1^N(Q^2)-\frac{Q^2}{4m_p^2}F_2^N(Q^2),\quad
G_M^N(Q^2)=F_1^N(Q^2)+F_2^N(Q^2), \quad N=p,n.
\end{displaymath}
   It has been known for some time that ChPT  results at ${\cal O}(q^4)$
only provide a decent description up to $Q^2=0.1\,\mbox{GeV}^2$ and
do not generate sufficient curvature for larger values of $Q^2$
\cite{Kubis:2000zd,Fuchs:2003ir}. To improve these results
higher-order contributions have to be included. This can be achieved
by performing a full calculation at ${\cal O}(q^5)$ which would also
include the analysis of two-loop diagrams.
   Another possibility is to include additional degrees of freedom, through which
some of the higher-order contributions are re-summed.
   Both the reformulated IR regularization and the EOMS scheme allow for
a consistent inclusion of vector mesons which already a long time
ago were established to play an important role in the description of
the nucleon form factors.
    Figure \ref{G_neu} shows the results for
the electric and magnetic Sachs form factors in the EOMS scheme
(solid lines) and the infrared renormalization (dashed lines)
\cite{Schindler:2005ke}.
   A {\em consistent} inclusion of vector
mesons clearly  improves the quality of the description.

\subsection{Axial and Induced Pseudoscalar Form Factors}
\begin{figure}[htb]
\begin{center}
\epsfig{file=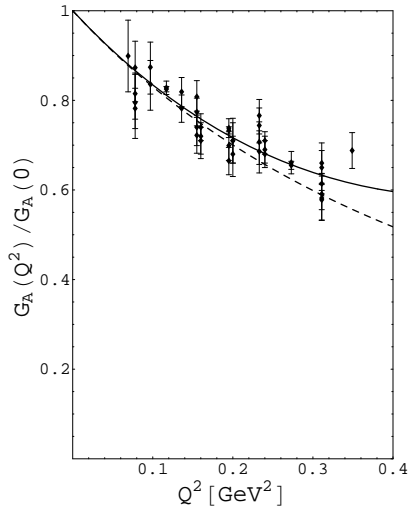,width=0.4\textwidth}
\end{center}
\caption{\label{H1_axff_GAwith} Axial form factor $G_A$ at ${\cal
O}(q^4)$ including the axial-vector meson $a_1$ explicitly. Full
line: result in infrared renormalization, dashed line: dipole
parametrization.}
\end{figure}

    Assuming isospin symmetry, the most general
parametrization of the isovector axial-vector current evaluated
between one-nucleon states is given by
\begin{displaymath}
\langle N(p')| A^{\mu,a}(0) |N(p) \rangle = \bar{u}(p')
\left[\gamma^\mu\gamma_5 G_A(Q^2) +\frac{q^\mu}{2m_N}\gamma_5
G_P(Q^2) \right] \frac{\tau^a}{2}u(p),
\end{displaymath}
where $q_\mu=p'_\mu-p_\mu$, $Q^2=-q^2$, and $m_N$ denotes the
nucleon mass.
   $G_A(Q^2)$ is the axial form factor and
$G_P(Q^2)$ is the induced pseudoscalar form factor.
     The value of the axial form factor at zero momentum transfer is defined as
the axial-vector coupling constant, $g_A=G_A(Q^2=0) =1.2695(29)$,
and is quite precisely determined from neutron beta decay. In Ref.\
\cite{Schindler:2006it} the form factors $G_A$ and $G_P$ have been
calculated in manifestly Lorentz-invariant baryon chiral
perturbation theory up to and including order ${\cal O}(q^4)$.
   In addition to the standard treatment including the
nucleon and pions, the axial-vector meson $a_1$ has also been
considered as an explicit degree of freedom.
   The inclusion of the axial-vector meson effectively results in one additional
low-energy coupling constant which has been determined by a fit to
the data for $G_A(Q^2)$.
   The inclusion of the axial-vector meson results in an improved
description of the experimental data for $G_A$ (see
Fig.~\ref{H1_axff_GAwith}).

\subsection{The Delta Resonance}
   A relativistic description of the spin-3/2 delta resonance
typically starts with the Rarita-Schwinger approach
\cite{Rarita:1941mf} in terms of a vector-spinor field $\Psi^\mu$.
   Such a description involves too many dynamical degrees of freedom
resulting in constraints.
   In the EFT program one needs to construct the most general
interaction Lagrangian.
   When dealing with systems involving constraints one has to make
sure that the corresponding Lagrangian equations of motion do not
involve an inconsistency.
   This is achieved with the aid of Dirac's analysis \cite{Dirac:2001}
using the Hamiltonian method.
   In a system with constraints there exist
certain relations connecting the momentum variables, of the type
$\Phi_m(q,p)=0$ which are referred to as primary constraints.
   Introducing the constraints in terms of Lagrange multipliers into
the Hamiltonian, $H_T=H+u_m\Phi_m$, one considers the time evolution
in terms of Poisson brackets, $\{H_T,\Phi_m\}=0$, thus generating
new (secondary) constraints.
   The procedure is iterated until all Lagrange multipliers have
been solved.
   In a {\em consistent} theory the number of initial degrees of
   freedom minus the number of constraints must equal the correct
   number of degrees of freedom.
   From this one obtains restrictions on the possible interaction
   terms.

For example, the $\mathcal{L}_{\pi\Delta}$ interaction Lagrangian
\cite{Hemmert:1997ye}
\begin{displaymath}
\mathcal{L}_{\pi\Delta}=
-\bar{\Psi}^{\mu}\biggl[\frac{g_1}{2}g_{\mu\nu}
\gamma^{\alpha}\gamma_5\partial_{\alpha}\phi
+\frac{g_2}{2}(\gamma_{\mu}\partial_{\nu}\phi
+\partial_{\mu}\phi\gamma_{\nu})\gamma_5+\frac{g_3}{2}\gamma_{\mu}
\gamma^{\alpha}\gamma_5\gamma_{\nu}\partial_{\alpha}\phi\biggr]\Psi^{\nu}
\end{displaymath}
contains three seemingly independent coupling constants.
   However, an analysis of the constraints yields \cite{Wies:2006rv}
\begin{displaymath}
g_2=Ag_1\,, \quad  g_3=-\frac{1}{2}(1+2A+3A^2)g_1,
\end{displaymath}
where $A$ is a parameter of the lowest-order Lagrangian.
   As a result of these constraints the total Lagrangian is
invariant under the so-called point transformation, guaranteeing
that the physical quantities are independent of the off-shell
parameter $A$.
   On the other hand, demanding the invariance under
the point transformation alone is less stringent and produces only a
class of relations among the coupling constants.
   The analysis of the constraint as a rule leads to a reduction in the number
of free parameters of the Lagrangian.

\section{Summary and Conclusions}
   Both the infrared regularization and the
EOMS scheme allow for a simple and consistent power counting in
manifestly Lorentz-invariant baryon chiral perturbation theory.
   The inclusion of vector and axial-vector mesons as explicit
degrees of freedom leads to an improved phenomenological description
of the electromagnetic and axial form factors, respectively.
   When dealing with the delta resonance, Dirac's constraint
   analysis leads to an identification of redundant parameters.

\begin{acknowledge}
   I would like to thank D.~Djukanovic, T.~Fuchs, J.~Gegelia,
C.~Hacker, G.~Japaridze, M.~R.~Schindler, and N.~Wies for the
fruitful collaboration on the topics of this talk. This work was
made possible by the financial support from the Deutsche
Forschungsgemeinschaft (SFB 443 and SCHE 459/2-1) and the EU
Integrated Infrastructure Initiative Hadron Physics Project
(contract number RII3-CT-2004-506078).
\end{acknowledge}

\end{document}